\documentclass[aps,prb,twocolumn,showpacs]{revtex4}  
\usepackage{epsfig}
\begin{document}
 \title{Edge spin accumulation in  2D electron and hole systems  in a  quasi-ballistic regime} 
 \author{Alexander Khaetskii \footnote{On leave from Institute of Microelectronics Technology, Russian Academy of Sciences, 142432 Chernogolovka, Moscow District, Russia; akhaetsk@buffalo.edu}}
\affiliation{Department of Physics, University at Buffalo, SUNY, Buffalo, NY 14260-1500} 

\date{\today}


\begin{abstract}
We consider a two-dimensional structure with spin-orbit-related  splitting of the electron (hole) spectrum and calculate the edge spin density which appears due to the intrinsic mechanism of spin-orbit interaction in the presence of a charge current through the structure.  We concentrate on the quasi-ballistic case when a mean free path, being much smaller than the sample size, is larger than the spin precession length determined  by the value of the spin-orbit splitting.  We show that regardless of the presence or absence of the bulk spin current the main source of the edge spin density is the boundary scattering itself.  The character of the edge spin density  depends on the smoothness of the bulk impurity  potential.  We have calculated the edge spin density profile 
for an {\it arbitrary} smoothness of the scattering potential in the bulk,  and discussed relation to the existing experiments for two-dimensional  holes. 
\end{abstract}


\maketitle

\section{Introduction}

  The spin Hall effect and edge spin accumulation  in two-dimensional (2D) structures  attracted recently a lot of attention \cite{Rashba,We}. 
Both phenomena are caused by spin-orbit coupling. 
There are two essentially different mechanisms of  the spin Hall effect, extrinsic \cite{Dyakonov} which is determined by the properties of impurities, and intrinsic one \cite{Sinova}, related to spin-orbit coupling in a perfect crystal and associated  splitting of the particle spectrum. 
The edge electron spin density accumulation, related to the Mott asymmetry in electron scattering  off impurities,  has been recently measured \cite{Kato}.  
  Moreover, the edge spin density in the  2D  hole system, which is due to the intrinsic mechanism of the spin-orbit interaction,  has also been observed \cite{Wunderlich}.  

  It is quite well established that in the diffusive regime (and when a spin diffusion length is much larger than a mean free path), 
the spin density appearing near the boundary is the result of the spin flux coming from the bulk  \cite{We,Dyakonov}.   In the particular case of the Rashba Hamiltonian it is still true, though some details  depend on the boundary conditions.   For the hard wall case,  when the spin current is zero at the boundary, 
 the spin current  and the spin density component perpendicular to the plane are zero everywhere down to the sample boundary \cite{BC}.
In the case when the boundary condition is the equality of a spin density to zero at the boundary,  a spin flux  is nonzero within spin precession length near the boundary and is precisely the source of the finite  $S_z$ component at the edge \cite{Bauer}.  This spin flux is directed  {\it towards} the boundary and  is caused by the electric field existing in the bulk.
 
   \par  Till recently  the opposite case, when the spin precession length $L_s=\hbar v_F/\Delta$ is much shorter than the mean-free path, has been studied mostly numerically.  (Here $v_F$ is the Fermi velocity,  and $\Delta$ is the spin-orbit-related splitting of the electron spectrum, which we consider to be smaller than the Fermi energy, $\Delta \ll E_F$). 
 It includes the case of finite size ballistic structures, when the mean-free path is much larger than the sample size (a mesoscopic spin Hall effect).  The existing literature for the ballistic case includes several papers \cite{Nikolic,Reynoso},  where the problem is treated numerically for the systems with the size which is comparable or less than the spin precession length.  
There are several analytical studies of the purely ballistic case \cite{Zyuzin,Silvestrov,Khaetskii1,Khaetskii2}. In particular, 
in Ref. \onlinecite{Khaetskii2} we found the characteristic scale for the edge  spin accumulation in the ballistic case  and studied how the scale depends on the boundary conditions.  It was also shown in Ref.\onlinecite{Khaetskii2} that the edge spin density in a mesoscopic spin Hall effect is not a result of the spin current flowing towards the boundary.

\par
In this paper we solve analytically the problem of edge spin accumulation  for a strong spin-orbit splitting  ($L_s \ll l$), but when the size of the structure is much larger than the mean free path $l=v_F \tau_p$, i.e. in the quasi-ballistic regime in our terminology ($\tau_p$ is a mean-free time).  
This case is very important in particular for the reason that it is related to already existing experimental data for 2D holes, see,  for example, Refs.(\onlinecite{Wunderlich,Numerics}).  
    The problem of the edge spin accumulation in the quasi-ballistic regime was considered before  both numerically and analytically for the linear Rashba Hamiltonian, see Refs.  [\onlinecite{Raimondi,Sonin}], and for 2D holes it was done numerically in  Ref. [\onlinecite{Numerics}]. (See the discussion of the results obtained in Ref.  [\onlinecite{Numerics}]  below). 
 Ref. [\onlinecite{Raimondi}] considers the problem numerically using the kinetic equation approach and only studies the case of smooth in transverse direction and straight boundary.  Therefore the boundary scattering does not induce transitions between sub-bands with opposite helicity, as a result  the authors do not observe any edge spin polarization, except in the corners of the structure. 
The problem has been considered analytically in Ref.\ [\onlinecite{Sonin}] for the Rashba Hamiltonian and for a short-ranged impurity potential in the bulk. The author concluded  that a smooth spin density profile with the scale $L_s$ appears near a boundary.  However, 
as it is shown in Ref. [\onlinecite{Khaetskii2}],  the proper treatment of the scattering problem in the bulk,  which determines  the correct form of the electron distribution functions, and the use of the unitarity of the boundary scattering \cite{Khaetskii2} leads to the cancellation of the smooth edge spin density component.  In Ref. [\onlinecite{Khaetskii2}] only the case of a $\delta$-correlated impurity potential was considered. 
\par

\begin{figure}
\begin{center}
\epsfig{file=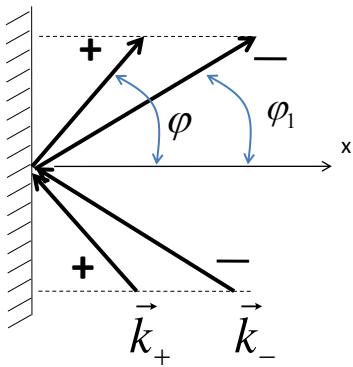,width=0.3\textwidth}
\end{center}
\caption{Schematics of the boundary specular scattering in the presence of spin-orbit coupling. Plus and minus modes are shown for the same energy and the same wave vectors along the boundary.} 
\label{fig:Spin1}
\end{figure}
    
In the quasi-ballistic regime the electric field  in the bulk of the sample is finite. Therefore,  the distribution functions for the electron states (i.e. the diagonal components of the spin-density matrix)  are determined by the electric field and by scattering off  the impurities in the bulk of a system \cite{Khaetskii}.  
 So far, there is no understanding of the mechanism of the edge spin accumulation in this regime.  In this case two sources of an edge spin accumulation are possible, the first one is due to the spin current coming from the bulk,  and the second one is a generation of the edge spin density upon the boundary scattering itself \cite{Zyuzin,Khaetskii2}. The relative role of those mechanisms has never been studied.  
 The bulk spin flux  is determined by the non-diagonal components of the spin-density matrix. However,  the edge source which is actually studied for the {\it ballistic} Rashba case in Refs. \cite{Zyuzin,Khaetskii2} is determined by the diagonal components of the spin density matrix. 
In the case of a strong spin-orbit splitting,  when $\Delta \tau_p/\hbar \gg 1$,   the contribution of the edge source can be larger since the diagonal components are proportional to the mean-free time, but the non-diagonal ones are not. 
\par
It is obvious, that the characteristic length near a boundary,  where the spin density arises, is the spin precession length.  Since this length is much smaller than a mean-free path, 
 the problem of the edge spin generation can be solved analytically with the use of the scattering states, see Fig.\ \ref{fig:Spin1}.   
The occupations of these states (the distribution functions, which are different from the ballistic case) are determined by the solution of the kinetic equations for the spin-density matrix in the bulk \cite{Khaetskii},  and are the input parameters for the part of the problem related to the scattering by the boundary.

\par It is important to understand how the phenomenon (for example, the edge spin density profile) depends on the character of the bulk scattering, i.e. the smoothness of the impurity potential, and the form of the spin-orbit Hamiltonian. Both questions  are answered in the present  paper. Besides the Rashba Hamiltonian we  also consider the cubic in 2D momentum spin-orbit Hamiltonian, i.e. 2D holes, the case which was probed experimentally in Refs. \onlinecite{Wunderlich,Numerics}.

\section{Rashba spin-orbit Hamiltonian}

       Rashba spin-orbit Hamiltonian  takes the following form
\begin{equation}
\hat {\cal H}({\bf p})=\frac{p^2}{2m}+ \frac{\alpha}{2}\vec{n}[\vec{\sigma}\times {\bf p}],
\label{Rashba}
\end{equation}
where $\vec{n}$ is the normal to the plane, $\vec{\sigma}$ are the Pauli matrices, and ${\bf p}$ is the 2D momentum. 
The solutions of this Hamiltonian corresponding to  the helicity values $ M=\pm $ have the form $\exp (i{\bf p}{\bf r}/\hbar)\chi_M({\bf p})$, where ${\bf r}=x,y$.  
The  spinors and corresponding eigenenergies are
$$\chi_{\pm}(\varphi)= \frac{1}{\sqrt{2}}\left( 
\begin{array}{ll}
1 & \\
\mp i e^{i\varphi} &
\end{array}
\right),  \,\,\,  \epsilon_M(p)= \frac{p^2}{2m}+ \frac{M}{2}\alpha p,$$
with $\varphi$ being the angle between the momentum ${\bf p}$ and the positive direction of the $x$-axis. 
\par
The scattering matrix for the case of scattering by an abrupt straight boundary (see Fig.\ \ref{fig:Spin1})  was found in Refs. \onlinecite{Khaetskii1,Khaetskii2}. 
Two scattering states, which obey zero boundary conditions 
at $x=0$, and which correspond to incident plus and minus modes with given wave vector along the boundary and the same energy are
\widetext
\begin{eqnarray}
\hat{\Psi}_{+}^{(0)}(x,y)=e^{ik_yy}[\chi_{+}(\pi -\varphi)e^{-ikx}+ F_{+}^{+}\chi_{+}(\varphi)e^{ikx}+               F_{+}^{-}\chi_{-}(\varphi_1)e^{ik_1x}]; \,\,\,  \hat{\Psi}_{+}^{(0)}(0,y)=0, 
\label{plusmode} \\
\hat{\Psi}_{-}^{(0)}(x,y)=e^{ik_yy}[\chi_{-}(\pi -\varphi_1)e^{-ik_1x}+ F_{-}^{+}\chi_{+}(\varphi)e^{ikx}+               F_{-}^{-}\chi_{-}(\varphi_1)e^{ik_1x}]; \,\,\,   \hat{\Psi}_{-}^{(0)}(0,y)=0.
\label{minusmode}
\end{eqnarray}
\endwidetext
 The wave vectors entering the above expressions are defined as follows
 \begin{equation}
 k^2=k_+^2-k_y^2, \,\,\, k_1^2=k_-^2-k_y^2, \,\,\, \hbar k_{\pm}=\hbar k_F \mp \frac{m\alpha}{2},
 \label{k_vectors}
 \end{equation}
 where $p_{\pm}=\hbar k_{\pm}$ are the momenta at the Fermi energy in the plus and minus modes, and $k_F$ is the Fermi wave vector in the absence of spin splitting. The angles $\varphi$, $\varphi_1$ are expressed as $\sin (\varphi)=k_y/k_+$ and $\sin (\varphi_1)= k_y/k_-$ (see Fig.\ \ref{fig:Spin1}).  
 The scattering amplitudes $F_{+}^{+}$ and  $F_{+}^{-}$ take the following form:
\begin{equation}
F_{+}^{+}=-\frac{(e^{i\varphi_1}-e^{-i\varphi})}{(e^{i\varphi_1}+e^{i\varphi})}; \,\,\, F_{+}^{-}= -\frac{2\cos\varphi}{(e^{i\varphi_1}+e^{i\varphi})}.
\label{ScattAmpl}
\end{equation}
 The amplitudes $F_{-}^{-}$ and  $F_{-}^{+}$ for the incident minus mode with the same $k_y$ and the same energy  are obtained from $F_{+}^{+}$ and  $F_{+}^{-}$ by replacing  $\varphi \leftrightarrow  \varphi_1$. Then for the components of the unitary scattering matrix $\hat{S}$ we obtain \cite{Khaetskii2}:
  \begin{equation}
  S_+^+=F_{+}^{+},\,\, S_-^-=F_{-}^{-}, \,\, S_+^-= S_-^+=F_{+}^{-}\sqrt{\frac{v_{x,-}}{v_{x,+}}},
  \label{SMatrix}
  \end{equation}
 where $v_{x,i}=\partial \epsilon_i/\partial p_x$ are the group velocities. In the case of Rashba Hamiltonian one has the relation
 $v_{x,-}/v_{x,+}= \cos\varphi_1/\cos\varphi$. 
 \par 

\subsection{The edge spin density}

 The average $z$-component of the spin as a function of coordinates is given by the following expression:
  \begin{eqnarray}
  \langle S_z(x)\rangle =\sum_{i=\pm}\int \frac{dk_y}{(2\pi)^2}\frac{d\epsilon}{v_{x,i}}f_i(\epsilon,k_y) \nonumber \\
 \times  \langle\hat{\Psi}_{i}^{(0)}(x)|\hat S_z |\hat{\Psi}_{i}^{(0)}(x)\rangle,
  \label{S_z}
\end{eqnarray}
 where $f_i(\epsilon,k_y)$ is the distribution function in the mode $i$ for given energy and wave vector $k_y$ along the boundary. 
\par
We use further the following notations: $f_+(\epsilon_F,k_y)=f_+(\vec{k}_+)$, $f_-(\epsilon_F,k_y)=f_-(\vec{k}_-)$.
These two functions are determined by bulk scattering in the presence of an electric field, and are not equal to each other in a general case, in contrast to the ballistic limit. In the case considered here the electric field is parallel to the y-axis, and the functions are presented in the form \cite{Khaetskii}
\begin{equation}
f_+(\vec{k}_+)\equiv f_+({k}_+)k_y/k_+,  \,\,\, f_-(\vec{k}_-)\equiv f_-({k}_-)k_y/k_-
\label{functions} 
\end{equation}

An interference of different terms in  Eqs. (\ref{plusmode}) and (\ref{minusmode}) gives rise to various contributions to $\langle S_z(x)\rangle $ with different oscillation periods. We are mostly interested  in this paper by  the smooth part  $\langle S_z(x)\rangle_s $  which involves the interference of the outgoing waves (two last terms in Eqs.\ (\ref{plusmode}) and (\ref{minusmode})):
  \begin{eqnarray}
\langle S_z(x) \rangle_s =2 {\rm Re}  \Bigg \{ \int \frac{dk_y}{(2\pi)^2}  \frac{d\epsilon}{\sqrt{v_{x,-}v_{x,+}}} 
\langle \chi_{-}(\varphi_1)|\hat{S}_z|\chi_{+}(\varphi) \rangle
\nonumber \\ 
\times
 e^{i(k-k_1)x} S_{+}^{+}\cdot (S_{+}^{-})^{\star}[ f_+(\epsilon,k_y)- f_-(\epsilon,k_y)  ] \Bigg \}. 
    \label{Unitary}
\end{eqnarray}
  Note, that the characteristic period of oscillations  in Eq.\ (\ref{Unitary}) is  the spin precession length. 
         
While deriving Eq.~(\ref{Unitary}) we used the unitarity condition $S_{+}^{+}\cdot (S_{+}^{-})^{\star}=-S_{-}^{+} \cdot (S_{-}^{-})^{\star} $, as a result the above expression  became proportional to the difference between the distribution functions $f_+(\epsilon,k_y)$ and $f_-(\epsilon,k_y)$. Note that in a purely ballistic case \cite{Khaetskii2} these functions are the Fermi functions of the leads with shifted chemical potentials, and for given $k_y$ (i.e. for given lead) these functions are equal to each other since they depend only on the energy. This was the reason for the absence of the smooth spin density near a straight boundary in a ballistic case \cite{Khaetskii2}. In the quasi-ballistic case considered here these functions are not equal to each other,  we will see that the relation between them depends on the nature of the scattering potential in the bulk. 

\subsection{Solution for the edge spin in the case of a short-ranged disorder potential in the bulk}
\par 
The kinetic equations for different components of the spin-density matrix in the quasi-ballistic regime ($L_s \ll l$) in the presence of an electric field and in the case of arbitrary spin-orbit Hamiltonian and arbitrary smoothness of the impurity potential treated in the Born approximation were derived in Ref. \cite{Khaetskii}.  Keeping only the diagonal components of the spin-density matrix in the collision term, from Eqs.(5,6) of Ref. \cite{Khaetskii} we obtain for the Rashba Hamiltonian in the regime $\Delta\gg \hbar/\tau_p$ the following exact relation between the functions $f_+({k}_+)$ and $f_-({k}_-)$ entering Eq. (\ref{functions})
 \begin{equation}
\frac{f_-({k}_-)}{f_+({k}_+)}=\frac{[a_{1,+}k_+ -\tilde{a}_2k_--k_+\tilde{a}_3]}{[a_{1,-}k_--\tilde{a}_2k_+-k_-\tilde{a}_3]}, 
  \label{relation}
  \end{equation}
 where 
\begin{eqnarray}
a_{1,\pm}= -\int \frac{d\theta}{2\pi}  \sin^2 \theta \tilde{W}(2k_{\pm}|\sin (\theta/2)|) , \,\, \nonumber \\
\tilde{a}_2= \int \frac{d\theta}{2\pi}  (1-\cos\theta) \tilde{W}(\sqrt{k_+^2+k_-^2 -2k_+k_-\cos\theta}), \,\, \nonumber \\
\tilde{a}_3= \int \frac{d\theta}{2\pi}\cos \theta (1-\cos\theta)\tilde{W}(\sqrt{k_+^2+k_-^2 -2k_+k_-\cos\theta}). 
\label{a}
\end{eqnarray}
Here the impurity scattering kernel ($\theta$ is the scattering angle) is
\begin{eqnarray} 
\tilde{W}(\theta)= \frac{n_i}{2\hbar^3}|U(\theta)|^2,\,\, U(\theta)=U(|\vec{k}_1-\vec{k}|)=  \nonumber \\
 U(\sqrt{k^2+k_1^2 - 2kk_1 \cos \theta}),  
\label{tilde_W}
\end{eqnarray}
where $n_i$ is the 2D impurity density, and $U(\vec{k} -\vec{k}_1)$ is the Fourier component of the impurity potential. 
To see how the accumulated spin density depends on the characteristic scale of the impurity potential, let us write down the condition of the equality of the 
distribution functions $f_+(\vec{k}_+)$ and $f_-(\vec{k}_-)$. From the definition of these functions, Eq.~(\ref{functions}),  and Eq.~(\ref{relation}) we obtain 
the following condition 
\begin{equation} 
(k_-^2-k_+^2)\tilde{a}_3=k_-^2 a_{1,-}-k_+^2a_{1,+}
\label{equality}
\end{equation}
First of all, we see that for the $\delta$-correlated potential when $\tilde{W}(\theta)$ does not depend on angle, the above Eq.~(\ref{equality}) is just an identity, i.e. the smooth component of the edge spin, Eq.~(\ref{Unitary}),  is absent.   The parameter which determines the magnitude of the effect is the ratio of the correlation radius of the impurity potential $d$ and the spin precession length $L_s$. Let us demonstrate that the above equality Eq.~(\ref{equality}) holds up to $\alpha^2$ order. In the first order we obtain
 \begin{equation} 
(k_F/2) da_1/dk_F=a_3-a_1, 
\label{first}
\end{equation} 
where $a_1, \,\, a_3$ are the corresponding quantities in Eqs.~(\ref{a}) calculated with the scattering kernel at the Fermi momentum: 
\begin{equation} 
\tilde{W}(2k_F|\sin (\theta/2)|) \equiv W(\theta)
\label{W}
\end{equation} 
The explicit form of Eq.~(\ref{first}) presented above is:
\begin{equation} 
-k_F\int_0^{\pi} d\theta \frac{dW(\theta)}{dk_F}\sin^2\theta =2\int_0^{\pi} d\theta W(\theta) [\cos\theta -\cos2\theta]
\label{integral}
\end{equation} 
Introducing a new variable $x=\sin(\theta/2)$, $0\leq x \leq 1$, we can express $dW(\theta)/dk_F=(x/k_F) dW/dx$, and integrate the left hand side of  Eq.~(\ref{integral}) over the $x$ by parts. Then we can easily see that the result is indeed equal to the expression standing in the right hand side. 
We note that Eq.~(\ref{first}) holds only under the condition $d\ll L_s$, when a quantity $\tilde{a}_3$ can be expanded in powers of $\alpha$. In the opposite limit, when  $d\gg L_s$,  the quantity $\tilde{a}_3$ can be neglected,  and Eq.~(\ref{equality}) is violated already in the first order in $\alpha$, see below. 
The  terms of the second order  in $\alpha$ are identically absent in Eq.~(\ref{equality}). 
Thus, we obtain that at $d\ll L_s$ the difference between the distribution functions $f_+(\epsilon_F,k_y)$ and $f_-(\epsilon_F,k_y)$ 
 is of the third order in $\alpha$: 
\begin{eqnarray} 
f_+(\epsilon_F,k_y)-f_-(\epsilon_F,k_y)=(\frac{m\alpha}{2p_F})^3  \Phi k_y \tilde{f}(\epsilon), \,\, \nonumber \\
\Phi=(a_1'+k_Fa_1''+k_F^2a_1'''/6-a_0')/a_1, 
\label{cubic}
\end{eqnarray} 
where $a_1'$, for example,  means the first derivative with respect to $k_F$, $a_1''$ the second derivative, etc. The quantity $\tilde{f}(\epsilon)=-(1/2ma_5) eE\partial f_0/\partial p_F$ depends only on the energy, and the two new coefficients are: 
\begin{eqnarray}
a_0= \int \frac{d\theta}{2\pi}  \cos \theta (1+\cos \theta)W(\theta), \,\, \nonumber \\
a_5= \int \frac{d\theta}{2\pi} (1-\cos\theta)W(\theta). 
\label{new_a}
\end{eqnarray}
Note, that $\tilde{f}(\epsilon)$ multiplied by $k_y/k_F$ has the meaning of the distribution function in the bulk in the presence of electric field $E$ for the spinless problem, $f_0$ is the Fermi distribution function,  and quantity $m a_5=1/\tau_p$  is the inverse transport scattering time. Thus,  the distribution function for the quasi-ballistic case is obtained from the one for a purely ballistic problem (with the bias $eV$ applied to the leads)  by  the following replacement 
$$
eV \rightarrow \hbar^2 k_y k_E/m, \,\,\, k_E=eE\tau_p/\hbar.
$$
Inserting result Eq.~(\ref{cubic}) in Eq.~(\ref{Unitary}), where all the quantities except  $f_+(\epsilon_F,k_y)-f_-(\epsilon_F,k_y)$ and  $e^{i(k-k_1)x}$ are calculated at $\alpha=0$, and doing a trivial integral over energy, we obtain: 
\begin{equation}
\langle S_z(x) \rangle_s = \frac{1}{(2L_s)^3} \frac{k_E}{2k_F} \Phi J(x), 
\label{result1}
\end{equation}
where 
$$
J(x)=k_F^{-3}\int_{-k_F}^{+k_F} \frac{dk_y k_y^2}{(2\pi)^2} \sin (k_1-k)x.
$$  
We can expand $k_1-k$ here with respect to the small parameter $ m\alpha/p_F$, and obtain 
$$
J(x)=\int_{0}^{1}\frac{dz z^2}{2\pi^2} \sin(\frac{x}{L_s\sqrt{1-z^2}}), \,\, z=k_y/k_F. 
$$
It is important that the characteristic values of $z$ which give the main contribution to the integral are of the order of unity, and the characteristic period of oscillations is $L_s=\hbar/m\alpha $.  

Integral $J(x)$ can be easily calculated in the limiting cases. At $x\ll L_s$ one has $J(x)=x/(8\pi L_s)$.  In opposite case $x\gg L_s$ we obtain $J(x)=(L_s/2\pi x)^{3/2}\cos [(x/L_s) +\pi/4]$. 

\subsection{Solution in the case of a smooth disorder potential}

In the limit of a smooth disorder potential,  $d\gg L_s$,  the quantity $\tilde{a}_3 $ is small, and we can neglect it. Then Eq.~(\ref{equality}) is violated already in the first order in $\alpha$. The corresponding difference of the distribution functions is:
\begin{eqnarray} 
f_+(\epsilon_F,k_y)-f_-(\epsilon_F,k_y)=(\frac{k_y}{k_F^2 L_s})  \Phi_1  \tilde{f}(\epsilon), \,\, \nonumber \\
\Phi_1=(2a_1+k_Fa_1')/(a_1-a_2).  
\label{linear}
\end{eqnarray} 
The result for the smooth spin density in this limit reads:
\begin{equation}
\langle S_z(x) \rangle_s = \frac{k_E}{2L_s}  \Phi_1 J(x). 
\label{result2}
\end{equation}
Note that the results Eq.~(\ref{result1}) and Eq.~(\ref{result2})
match at $d\simeq L_s$. Indeed, since in this parameter range scattering is of the low-angle type (we consider the case $L_s \gg \lambda_F$) with the characteristic scattering angle $1/(k_F d) \ll 1$, we can estimate the quantities $\Phi$ and $\Phi_1$ as follows: 
$\Phi \simeq d^2k_F$, $\Phi_1 \simeq 1$. While estimating the $\Phi$ value, 
we took into account the fact that for a small-angle scattering $(a_0/a_1) \simeq (k_Fd)^2 \gg 1$.  
Using  these values of $\Phi$ and $\Phi_1$, we obtain that results Eq.~(\ref{result1}) and Eq.~(\ref{result2})  match at $d\simeq L_s$.

\par
Note that the function $\Phi$ at arbitrary value of $d<L_s$ (including the case of a short-ranged scattering $d\ll \lambda_F$)
can be estimated as $\Phi \simeq d^2 k_F$. Thus, the results for the spin density obtained above,  Eqs.(\ref{result1},\ref{result2}), can be presented in the form 
\begin{equation}
\langle S_z \rangle_s \simeq \frac{k_E}{L_s}\frac{d^2}{L_s^2} \,\,\, at \,\, d<L_s; \,\, \langle S_z \rangle_s \simeq \frac{k_E}{L_s} \,\,\, at \,\, d>L_s. 
\label{main}
\end{equation}

\subsection{Fast oscillating contribution to the edge spin}

\par We note also that besides the smooth spin density component,  there is also a fast oscillating contribution with $2k_F$ wave vector,  in a complete analogy with the ballistic case \cite{Zyuzin,Khaetskii1,Khaetskii2}. If one takes in Eq.~(\ref{Unitary}) the distribution functions $f_+(\epsilon_F,k_y), \, f_-(\epsilon_F,k_y)$ in zeroth order with respect to $\alpha$ (which are equal to each other), then the smooth contribution  to the spin density with the scale $L_s$ is zero.  The  remaining fast contribution which is 
due to the interference between incoming and the outgoing waves [for example, between first and second terms in Eq.\ (\ref{plusmode})], and additional  contribution from the evanescent modes, see Ref.\onlinecite{Khaetskii2},  can be written in the form $\langle S_z(x)\rangle_f= (\hbar^2 k_E /m) (1/8\pi^2mv_F^2){\rm Im} I(x) $, where
\begin{equation}
I(x)=\int _{0}^{k_-}dk_y[k_+k_-  + k_y^2 -kk_1](e^{ikx}-e^{ik_1x})^2.
\label{fast}
\end{equation}
Once again, this contribution is obtained from the corresponding one in Ref.\onlinecite{Khaetskii2} just by the replacement 
$eV \rightarrow (\hbar^2 k_y k_E/m)$, which was indicated above. Further consideration is exactly the same as in Ref. \cite{Khaetskii2}, see Fig. 2 there.  The integrand function in $I$ is an analytical function of the variable $k_y$ in the right half plane ${\rm Re} k_y> 0$ (for positive $x$), and we can transform the original contour into the one shown in Fig. 2, Ref.\onlinecite{Khaetskii2}.   The power law decaying part of the integral $I$  comes from the imaginary axis of $k_y=i\kappa$.  Then, for $x\gg \lambda_F$  the latter integral is determined by small $\kappa\ll k_F$, and for the case $\lambda_F \ll x \ll L_s^2 k_F$ one has  $ {\rm Im} I\approx -2\sqrt{\pi} (k_F/ x)^{3/2} \sin^2(x/2L_s)   \sin(2k_Fx+\pi/4)$. 
Note that  the total spin per unit length along the boundary calculated with the function Eq. \ref{fast} is given by
 $\int_0^{\infty}dx \langle S_z(x)\rangle_f  \simeq   k_E /(k_F^2 L_s^2)$. The main contribution to this integral comes from small distances from the boundary, $x\simeq\lambda_F$.
\par
Total spin calculated with the smooth part, see Eq.(\ref{main}), for the case $d<L_s$ reads
 $\int_0^{\infty}dx \langle S_z(x)\rangle_s  \simeq k_E d^2 /L_s^2$. Then the total contribution from the smooth part is larger than the corresponding contribution from the fast part if the condition $d >\lambda_F$ is fulfilled.  

\section{2D holes}

\par We note that in the case of general spin-orbit Hamiltonian with the spectrum 
\begin{equation}
\epsilon_M(p)= \frac{p^2}{2m}+ M\alpha p^N, \,\,\,  M=\pm 1/2
\end{equation}
 the functions which we need,  $f_-(\vec{k}_-)$ and  $f_+(\vec{k}_+)$,   are not equal to each other,   and their difference is of the first order in $ \alpha$ for an arbitrary scattering potential in the bulk,  including $\delta$-potential (at $N\neq 1$).  For the latter case 
a solution of bulk kinetic equations gives  the following relation between the distribution functions introduced above $f_-(k_-)v_+(k_+)=f_+(k_+)v_-(k_-)$, where  $v_+(k_+)=\hbar k_F/m+ \tilde \alpha (N-1)/2 $ and  $v_-(k_-)=\hbar k_F/m- \tilde \alpha (N-1)/2 $ are the corresponding velocities calculated at $k_+$ and $k_-$ in the linear with respect to $\tilde \alpha$ approximation, and $\tilde \alpha =\alpha p_F^{N-1}$.  Then we obtain for $N=3$ case, which corresponds to 2D holes 
$$
 f_+(\epsilon_F,k_y)-f_-(\epsilon_F,k_y)=(\frac{3 k_y}{L_s k_F^2})    \tilde{f}(\epsilon)
$$
 The result for the smooth edge spin density in the case of  $\delta$-correlated impurity potential reads \cite{amplitudes}:
\begin{equation}
\langle S_z(x) \rangle_s = \frac{3k_E}{2L_s} J_1(x). 
\label{result3}
\end{equation}
Here $L_s=\hbar/m \tilde\alpha$ and 
$$
J_1(x)=\int_{0}^{1}\frac{dz z^2}{2\pi^2} [3+16z^2(z^2-1)] \sin(\frac{x}{L_s\sqrt{1-z^2}}) 
$$
At $x\ll L_s$ one has $J_1(x)=x/(8\pi L_s)$.  In opposite case $x\gg L_s$ we obtain $J_1(x)=3(L_s/2\pi x)^{3/2}\cos [(x/L_s) +\pi/4]$. 
The degree of edge spin polarization which follows from Eq. (\ref{result3}) is
$$
\eta \simeq \frac{\Delta}{E_F}\frac{k_E}{k_F}
$$

\par 
We would like to mention that experiments Refs. (\onlinecite{Wunderlich,Numerics}) have dealt with  the edge spin accumulation in 2D hole system,  
where the observed effect  
 was interpreted as being caused by the intrinsic mechanism related to the cubic splitting of the energy spectrum. The parameter $\Delta \tau_p/\hbar$ in the numerical calculations was of the order of unity, the maximal value was only about four.  Moreover, the parameter $\Delta/E_F =0.4$, therefore the values of $\lambda_F/2$ and $L_s$ were close to each other. That is why a direct comparison of the results obtained here and in Refs. (\onlinecite{Wunderlich,Numerics}) is  difficult to make. 
It is also true  because of the large experimental uncertainty of the observed spin density value. 
In particular, it is not clear at all what was the characteristic spatial scale of the edge spin accumulation in the experiment. 
It seems that the authors of Ref. (\onlinecite{Numerics}) obtained numerically the edge spin density oscillating with $2k_F$. 
It remains unclear why the spin density calculated by them does not depend on $\tau_p$.  The authors concluded that the edge spin density is brought by the spin flux coming from the bulk of the sample.

We want to mention here,  however, that    in the quasi-ballistic limit $\Delta \tau_p/\hbar \gg 1$ the spin density which is due to the boundary scattering itself is parametrically larger than the one due to the incoming spin flux from the bulk. Indeed, the first contribution,  Eq. (\ref{result3}), has the value $eE\tau_p/\hbar L_s$ and the second one is $q_{xz}/v_F \simeq eE/\hbar v_F$. Their ratio is $\Delta \tau_p/\hbar \gg 1$, where we have used $L_s=\hbar v_F/\Delta$.  
\par 
In conclusion, we have solved analytically the problem of the edge spin accumulation which appears in the presence of a charge current  in 2D structures and is related to the intrinsic mechanism of spin-orbit interaction.  We concentrate on the quasi-ballistic case when the spin-orbit-related  splitting of the electron (hole) spectrum $\Delta$ is larger than the broadening $\hbar/\tau_p$ of the spectrum due to impurity scattering in the bulk of the structure. Otherwise the sample is in the diffusive regime, i.e. the sample sizes are much larger than a mean-free path. We have calculated the smooth component of the edge spin density which is located over the characteristic scale near the boundary determined by the spin precession length $L_s$.  This contribution appears due to  boundary scattering itself and in the case $\Delta \tau_p/\hbar \gg 1$ is larger than the contribution brought by the intrinsic spin flux from the bulk. 
 The precise characteristics  of the edge spin density profile  depend on the smoothness of the bulk impurity  potential.  We have discussed relation of the obtained results to the existing experiments for 2D holes.

 \par 
 \section{Acknowledgments}

I acknowledge the financial support from the  SPINMET project (FP7-PEOPLE-2009-IRSES).  I am also grateful to the Kavli ITP, Santa Barbara for the hospitality and to all participants of the Workshop "Spintronics: Progress in Theory, Materials, and Devices" for useful discussions.


\begin{thebibliography}{99}


\bibitem{Rashba} 
H.A. Engel, E.I. Rashba, and B.I. Halperin, in {\it Handbook of Magnetism and Advanced Magnetic Materials}, ed. by H. Kronm{\"u}ller and S. Parkin, Vol.5 (John Wiley and Sons, New York, 2007). 

\bibitem{We} 
M.I. Dyakonov, and A.V. Khaetskii, in {\it Spin Physics in Semiconductors}, Springer Series in Solid- State Sciences, ed. by M.I. Dyakonov (Springer, Berlin, 2008). 

\bibitem{Dyakonov}
M. I. Dyakonov and V.I. Perel, JETP Lett. {\bf 13}, 467 (1971); 
M. I. Dyakonov and V.I. Perel, Phys.  Lett. A {\bf 35}, 459 (1971). 
 
\bibitem{Sinova} 
S. Murakami et al., Science {\bf 301}, 1348 (2003); J. Sinova et al., Phys. Rev. Lett.  {\bf 92}, 126603 (2004). 


\bibitem{Kato}
Y.K.Kato et al., Science {\bf 306}, 1910 (2004). 


\bibitem{Wunderlich} 
J. Wunderlich et al.,  Phys. Rev. Lett.  {\bf 94}, 047204 (2005). 

\bibitem{BC}
Ya. Tserkovnyak  et al., Phys. Rev. B {\bf 76}, 085319 (2007); O. Bleibaum, Phys. Rev. B {\bf 74}, 113309 (2006). 

\bibitem{Bauer}
I. Adagideli and G.E.W. Bauer,  Phys. Rev. Lett.  {\bf 95}, 256602 (2005). 

\bibitem{Nikolic} 
B.K. Nikoli\'c et al.,   Phys. Rev. Lett. {\bf 95}, 046601 (2005);
 B.K. Nikoli\'c, L.P. Z{\^a}rbo, and S. Souma, Phys. Rev. B {\bf 73}, 075303 (2006). 

\bibitem{Reynoso}
G. Usaj and C.A. Balseiro, Europhys. Lett. {\bf 72}, 621 (2005);
A. Reynoso et al., Phys. Rev. B {\bf 73}, 115342 (2006).


\bibitem{Zyuzin} 
V.A. Zyuzin, P.G. Silvestrov, and E.G. Mishchenko,  Phys. Rev. Lett.  {\bf 99}, 106601 (2007). 

\bibitem{Silvestrov}
P.G. Silvestrov, V.A. Zyuzin, and E.G. Mishchenko,  Phys. Rev. Lett.  {\bf 102}, 196802 (2009). 

\bibitem{Khaetskii1}
 A. Khaetskii, E. Sukhorukov,   JETP  Letters {\bf 92}, 244 (2010). 

\bibitem{Khaetskii2}
A. Khaetskii, E. Sukhorukov, Phys. Rev. B {\bf 87}, 075303  (2013).


\bibitem{Numerics}
K. Nomura et al., Phys. Rev. B {\bf 72}, 245330 (2005).

 \bibitem{Raimondi} 
Raimondi et al., Phys. Rev. B {\bf 74}, 035340 (2006). 

\bibitem{Sonin}
E.B.  Sonin,  Phys. Rev. B {\bf 81},  113304  (2010). 

\bibitem{Khaetskii}
A. Khaetskii,  Phys. Rev. B {\bf 73}, 115323  (2006). 


\bibitem{Note}
  Note the different meaning of quantities $k_+$ and $k_-$ used here, and quantities $p_+$ and $ p_-$ used in \cite{Khaetskii}. 



\bibitem{amplitudes}
We calculated the result in the linear in $\tilde \alpha$
 approximation. It means that the scattering  amplitudes by the boundary are calculated at $\tilde \alpha =0$. Then even for the hard wall conditions the troubles specific for the cubic Hamiltonian, see Ref. [\onlinecite{Khaetskii2}], do not show up, and the unitarity of scattering is fulfilled   for "+"
 and "-" modes with $S_+^+=\sin 3\varphi$, $S_+^-=\cos 3\varphi$. 



\end{thebibliography}
\end{document}